\documentclass[twocolumn]{aastex63}
\newcommand{\meth}{CH$_3$OH}

\newcommand{\kms}{km~s$^{-1}$}
\newcommand{\jybeam}{Jy~beam$^{-1}$}
\newcommand{\Blos}{\ensuremath{B_{\rm los}}}
\shorttitle{The Zeeman Effect in 38~GHz Class~II methanol masers}
\shortauthors{Momjian \& Sarma}

\begin{document}

\title{The Discovery of the Zeeman Effect in 38~GHz Class~II Methanol Masers}


\correspondingauthor{E.\ Momjian}
\email{emomjian@nrao.edu}

\author{E. Momjian}
\affiliation{National Radio Astronomy Observatory \\
P. O. Box O, Socorro, NM, USA}

\author{A. P. Sarma}
\affiliation{Department of Physics \& Astrophysics, DePaul University \\
2219 N. Kenmore Ave, Byrne 211, Chicago IL 60614, USA\\
\textit{Received 2023 August 8; revised 2023 September 6; accepted 2023 September 26, for publication in ApJ}
}

\begin{abstract}

Magnetic fields likely play an important role in star formation, but the number of directly measured magnetic field strengths remains scarce. We observed the 38.3 and 38.5~GHz Class~II methanol (\meth) maser lines toward the high mass star forming region NGC~6334\,F for the Zeeman effect. The observed spectral profiles have two prominent velocity features which can be further decomposed through Gaussian component fitting. In several of these fitted Gaussian components we find significant Zeeman detections, with $z\Blos$ in the range from 8 to 46~Hz. If the Zeeman splitting factor $z$ for the 38~GHz transitions is of the order of $\sim$1~Hz mG$^{-1}$, similar to that for several other \meth\ maser lines, then magnetic fields in the regions traced by these masers would be in the range of 8-46~mG. Such magnetic field values in high mass star forming regions agree with those detected in the better-known 6.7~GHz Class~II \meth\ maser line. Since Class~II \meth\ masers are radiatively pumped close to the protostar and likely occur in the accretion disk or the interface between the disk and outflow regions, such fields likely have significant impact on the dynamics of these disks.

\end{abstract}


\section{Introduction} \label{sec:intro}

A complete understanding of the role of magnetic fields in star formation remains elusive, even though major strides have been made in understanding how magnetic fields impact the star formation process (\citealt{pattle+2023}; \citealt{tsukamoto+2023}). Masers offer the opportunity to observe star forming regions at high angular resolution, because they are bright and compact sources \citep{richards+2020}. In particular, Class~II methanol (\meth) masers are known to form close to protostars, pumped by infrared radiation from the protostar itself. \citet{ellingsen+2018} observed 38.3 and 38.5~GHz Class~II \meth\ maser transitions at high angular resolution ($2\rlap{\arcsec}{.}\,3 \times 1\rlap{\arcsec}{.}\,4$) with the Australia Telescope Compact Array (ATCA). One of the objects in their sample, NGC~6334\,F, shows strong maser lines ($> 100$\,Jy) at these frequencies. Consequently, this prompted us to target NGC~6334\,F for the Zeeman effect, which is the most direct method for measuring magnetic fields in star forming regions.

There are two classes of \meth\ masers: Class~I and Class~II (\citealt{menten+1991}). Class~I \meth\ masers in star forming regions are known to be collisionally pumped in outflows (see, e.g., \citealt{leurini+2016}). Class~II \meth\ masers, meanwhile, are pumped by infrared radiation and are therefore located close to protostars (\citealt{cragg+2005}). They are exclusively associated with high mass star forming regions (\citealt{ellingsen+2006}). The most observed and well known Class~II transitions are at 6.7~GHz and 12.2~GHz (see, e.g., \citealt{nguyen+2022}, and references therein), but about 20 different Class~II \meth\ maser transitions have been observed to date (\citealt{breen+2019}, and references therein). Maser emission in the Class~II \meth\ lines at 38.3 and 38.5~GHz was discovered by \citet{haschick+1989} through single-dish observations. \citet{ellingsen+2018} carried out radio interferometric observations
of the 38.3 and 38.5~GHz methanol masers toward a sample of 11 high mass star forming regions that host strong 6.7~GHz \meth\ masers and detected 38.3~GHz transitions toward seven sources in their sample, and 38.5~GHz transitions toward six sources. They found that these transitions arise from the same location as the strong 6.7~GHz Class~II \meth\ maser transitions, although they are less co-spatial with the 6.7~GHz maser spots compared to the 12.2~GHz masers.

\begin{figure}[htb!]
\epsscale{1.15}
\plotone{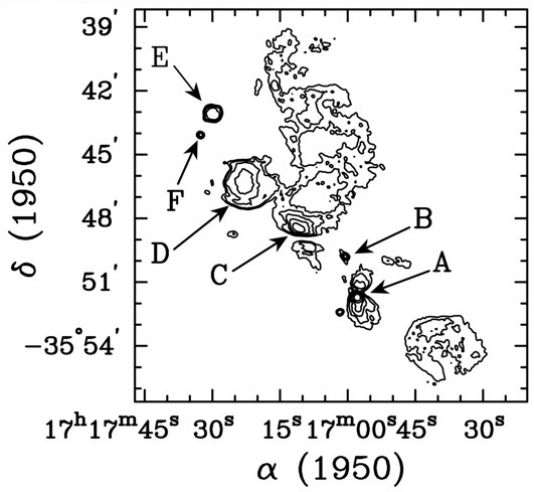}
\caption{Contour image of the 1.67~GHz continuum from NGC~6334 taken from \citet{sarma+2000}, showing the sources A-F (\citealt{rcm+82}). The observations reported in this work are toward source~F.\label{fig:s2000} }
\end{figure}

NGC~6334 is a well-known molecular cloud complex and star-forming region located at a distance of 1.3~kpc (\citealt{chibueze+2014}). Strong infrared and mm sources, hypercompact (HC) and ultracompact (UC) H~II regions, and powerful outflows reveal evidence of ongoing high mass star formation activity in this region (\citealt{rodriguez+2007}; \citealt{andre+2016}; \citealt{brogan+2016}; \citealt{hunter+2021}).
At 4.9~GHz, \citet{rcm+82} found six discrete continuum sources in NGC 6334. These six sources lie along a ridge of emission parallel to  the Galactic Plane and were named A-F in order of increasing Right Ascension (R.A.), as shown in Figure~\ref{fig:s2000}. Source~F, the target of the observations reported in this paper, is highly obscured at optical wavelengths, but prominent in cold dust traced by 850~$\mu$m submm continuum observations (\citealt{matthews+2008}). The dust clump associated with source~F 
was determined by \citet{matthews+2008} to have a mass of $2000~M_\odot$, indicating that it is a large reservoir of material for star formation. Also coincident with source~F is the far-infrared (FIR) source~I (\citealt{emerson+1973}; \citealt{loughran+1986}). In the CO $J=3-2$ transition, \citet{zhang+2014} observed a blueshifted outflow toward the southwest of the FIR source~I, and a redshifted outflow toward the northeast. \citet{hunter+2006} discovered four 1.3~mm continuum sources toward NGC~6334~I; these were later renamed as MM1-MM4 by \citet{brogan+2016}. Of these, the ultracompact (UC) H~II region MM3 overlaps with the peak of NGC~6334\,F. An outburst in the mm emission from MM1, accompanied by simultaneous flaring in multiple maser species, was ascribed to episodic accretion in this source (\citealt{hunter+2021}, and references therein).

In this paper, we report the detection of the Zeeman effect in the 38.3~GHz and 38.5~GHz Class~II \meth\ masers toward NGC~6334\,F. The details of the 2017 observations and data reduction are given in \S~\ref{sec:obsredn}. The results are presented in \S~\ref{sec:res}, and discussed in \S~\ref{sec:disc}. In \S~\ref{sec:conc}, we state our conclusions.

\section{Observations and Data Reduction} \label{sec:obsredn}

The observations of the Class II \meth\ maser emission lines
$6_2 \rightarrow 5_3$~A$^{-}$ at 38.3 GHz and $6_2 \rightarrow 5_3$~A$^{+}$ at 38.5 GHz
toward NGC~6334\,F were carried out with the Karl G. Jansky Very Large Array
(VLA)\footnote{The National Radio Astronomy Observatory (NRAO) is a facility of the
National Science Foundation, operated under cooperative agreement by Associated Universities, Inc.}
on 2021 March 23 and 2021 April 7. Each of these observing sessions was 2 hr long. The VLA was in its most compact (D) configuration with a maximum baseline length of 1~km.

\begin{deluxetable*}{lccccccrrrrrrrrcrl}
\tablenum{1}
\tablewidth{0pt}
\tablecaption{Parameters for VLA Observations
	\protect\label{tOP}}
\tablehead{
\colhead{Parameter} & 
\colhead{38.3 GHz Value} & \colhead{38.5 GHz Value} }
\startdata
Date \dotfill & \multicolumn{2}{c}{2021 Mar 23 \& Apr 7}     \\  
Configuration \dotfill & \multicolumn{2}{c}{D} \\
R.A.~of field center (J2000) \dotfill & \multicolumn{2}{c}{17$^{\text{h}}$~20$^{\text{m}}$~53$\fs$370} \\
Dec.~of field center (J2000) \dotfill & \multicolumn{2}{c}{$-$35\arcdeg~47\arcmin~02\rlap{\arcsec}.\,0} \\
Total bandwidth (MHz) \dotfill & \multicolumn{2}{c}{4}  \\
No.~of channels \dotfill & \multicolumn{2}{c}{1024} \\
Channel spacing (km~s$^{-1}$) \dotfill & \multicolumn{2}{c}{0.0305} \\
Approx.~time on source (hr) \ldots & \multicolumn{2}{c}{2.83} \\ 
Rest frequency (GHz) \dotfill  & 38.293270 & 38.452629 \\
FWHM of synthesized beam \dotfill & $5\, \rlap{\arcsec}.\, 33 \times 1\, \rlap{\arcsec}.\, 53$ 
							& $5\, \rlap{\arcsec}.\, 60 \times 1\, \rlap{\arcsec}.\, 50$ \\
& P.A.\, = 6.99\arcdeg & P.A.\, = 6.63\arcdeg \\ 
Line rms noise (mJy~beam$^{-1}$) \tablenotemark{a} & 10.5 & 12.0 \\
\enddata
\tablenotetext{a}{The line rms noise was measured from the Stokes $I$ image cube using maser line free channels.}
\end{deluxetable*}

The Wideband Interferometric Digital ARchitecture (WIDAR) correlator was configured to deliver 4~MHz sub-bands with dual polarization products (RR, LL) and 1024 spectral channels per sub-band. The resulting channel spacing was 3.90625~kHz, which corresponds to $\sim 0.0305$~km~s$^{-1}$ at the observed frequencies. In addition to NGC~6334\,F, the source J1331+3030 (3C 286) was observed as the absolute flux density scale calibrator. The uncertainty in the flux density calibration at the observed frequencies, accounting for various observational parameters (e.g., weather, reference pointing, and elevation
effects), is expected to be up to 10\%. We also observed the source J1720$-$3552 as the complex gain calibrator for part of each observing session to derive the absolute position of the masers in the target source. The phase referencing cycle time was six minutes; 5 minutes on the target and 1 minute on the complex gain calibrator. All the data reduction steps, including calibration, imaging, and deconvolution, were carried out independently for each observing session using the Astronomical Image Processing System (AIPS; \citealt{greisen+1990}). After Doppler correcting the spectral-line data of each transition and observing session independently, the spectral channel with the brightest maser emission in each was split off, and self-calibrated first in phase, then in both phase and amplitude, and imaged in iterative cycles. The final self-calibration solutions were then applied to the corresponding full spectral-line $uv$ data sets of NGC~6334\,F from each session. We note that AIPS calculates the Stokes parameter $I$
as the average of the right circular polarization (RCP) and left circular polarization (LCP), so that $I$ = (RCP + LCP)/2, whereas Stokes V is calculated by AIPS as half the difference between RCP and LCP, so that $V$ = (RCP $-$ LCP)/2; henceforth, all values of $I$ and $V$ are based on this implementation in AIPS. Also, we note that RCP is defined here in the standard radio convention, in which it is the clockwise rotation of the electric vector when viewed along the direction of wave propagation. Table 1 summarizes the parameters of the VLA observations and the corresponding synthesized beamwidths at full width half maximum (FWHM) and other parameters for each observing session.

\section{Results} \label{sec:res}

\begin{figure*}[htb!]
\epsscale{0.8}
\plotone{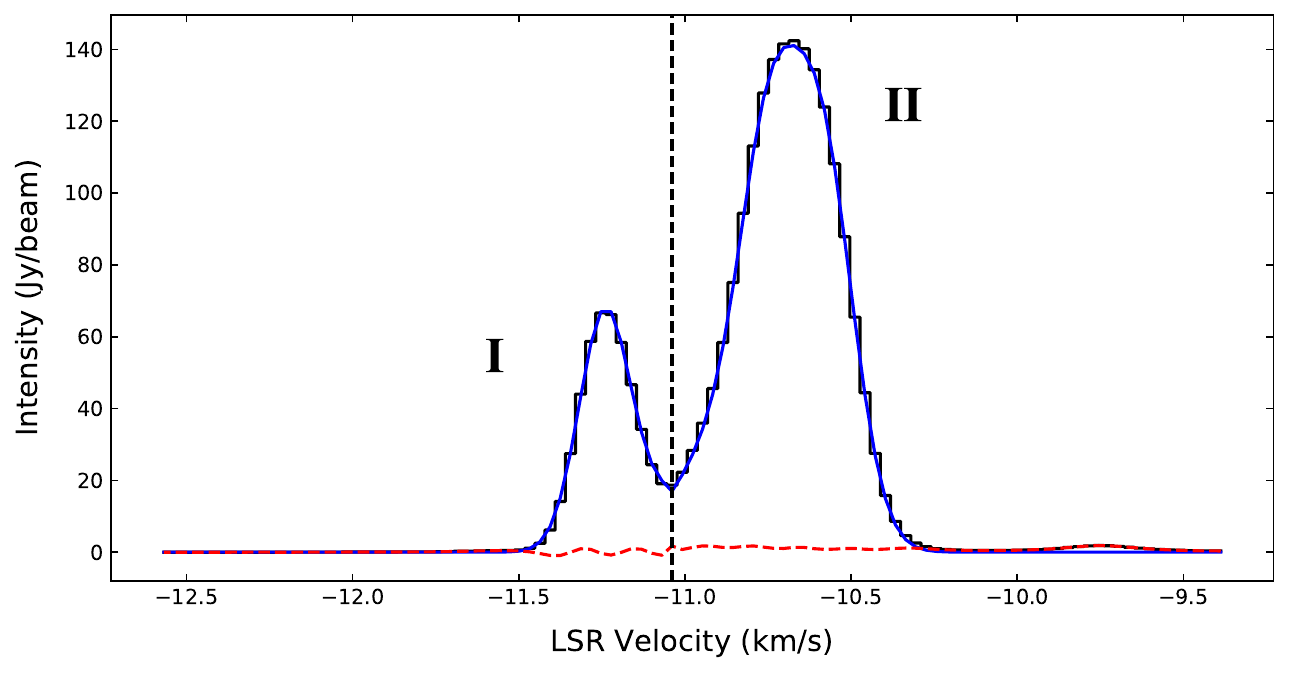}
\caption{Observed spectral profile (black histogram-like line) toward NGC 6334~F at 38.3 GHz showing two prominent spectral features in velocity, which we have labeled as I and II. The dashed vertical line separates these two features. The blue curve to the left of the dashed vertical line is the resultant profile obtained by summing components I and I-s listed in Table~\ref{tline} and shown in Figure~\ref{fig:IF2wG}, and the blue curve to the right of the dashed vertical line is the resultant profile obtained by summing components IIa, IIb, and IIc. The red dashed curve represents the residuals from the fit. \label{fig:IF2} }
\end{figure*}

\begin{figure*}[ht!]
\epsscale{0.8}
\plotone{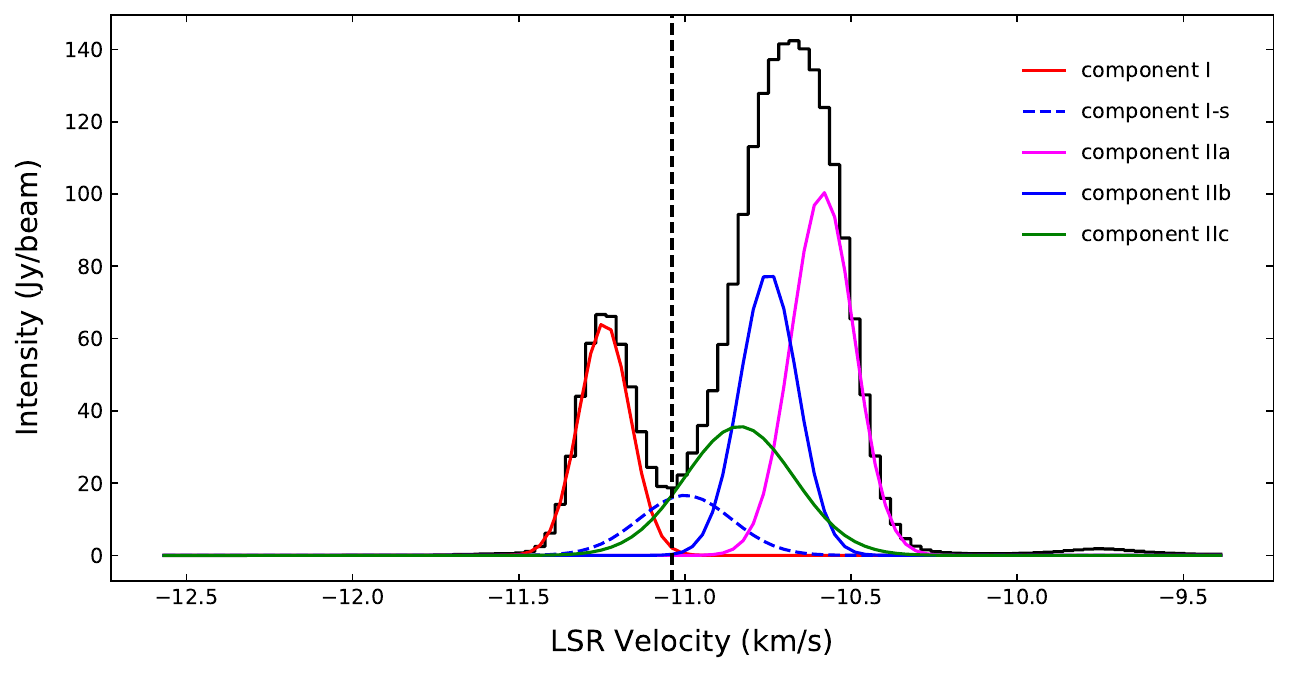}
\caption{Same spectral profile (black histogram-like line) as in Figure~\ref{fig:IF2}, but now showing the individual Gaussian components fitted to the observed profile at 38.3~GHz. The solid red and dashed blue curves show components~I and I-s listed in Table~\ref{tline}. The magenta, solid blue, and green curves show components IIa, IIb, and IIc respectively. Note that for visual clarity, the entire dashed blue and solid green curves are shown, even though the dashed blue curve is used only in the fit for spectral feature~I shown in Figure~\ref{fig:IF2}, and the solid green curve is used only in the fit for feature~II. \label{fig:IF2wG} }
\end{figure*}

We detected Class~II \meth\ masers at 38.3 and 38.5~GHz toward the position located at R.A. (J2000) = 17$^{\text{h}}$~20$^{\text{m}}$~53$\fs$37, Decl. (J2000) = $-$35\arcdeg~47\arcmin~01\rlap{\arcsec}.\,40 in NGC~6334\,F. The spectral profile of the Class~II methanol maser line detected at 38.3~GHz toward this position is shown in Figure~\ref{fig:IF2}. It has two prominent spectral features in velocity. We have labeled them as I and II in the figure, and separated them by a vertical dashed line. 
Visually, spectral feature~I looks like it could be fit by a single Gaussian component, but it did require a second shallower and broader component to fit the line wing on the side nearer to $-11$~km/s. We have labeled these as component~I and I-s respectively; despite the risk of confusion, we have chosen this nomenclature to emphasize that the spectral feature~I in Figure~\ref{fig:IF2} is almost a single-component Gaussian. The peak intensity of components~I and I-s, together with the velocity at line center and the Full Width at Half Maximum (FWHM) linewidth, are listed in Table~\ref{tline} and shown in Figure~\ref{fig:IF2wG}. Henceforth, we have tried to be clear from the context whether we are referring to spectral feature~I shown in Figure~\ref{fig:IF2}, or component~I that is listed in Table~\ref{tline} and shown in Figure~\ref{fig:IF2wG}. The peak intensity of component~I listed in Table~\ref{tline} is 64.49~\jybeam\ and its center velocity is $-11.24$~\kms. It is narrow, with a linewidth of 0.18~\kms. Component I-s is shallower, with about 1/4 the peak intensity of component~I. It is also broader, with a FWHM linewidth of 0.327~\kms. Note that even though component~I-s extends into the velocity space of component~II, it was not used in the fit for component~II. 
Meanwhile, at 38.5~GHz we observe a spectral profile that resembles the 38.3~GHz profile. As listed in Table~\ref{tline}, the intensity of component~I at 38.5~GHz is similar to that at 38.3~GHz; likewise for component~I-s. Within the errors, the center velocities and FWHM linewidths at 38.3 and 38.5~GHz are the same for components I and I-s.

\begin{deluxetable*}{ccccccccccccl}
\tablenum{2}
\tablecaption{Maser Line Parameters from Gaussian Fits \label{tline} } 
\tablehead{
	\colhead{Frequency} && \colhead{} && \colhead{Intensity} && \colhead{Center Velocity} && \colhead{Velocity Linewidth\tablenotemark{{\rm \scriptsize a}}} \\
	\colhead{(GHz)} && \colhead{Component} && \colhead{(\jybeam)} && \colhead{(\kms)} && \colhead{(\kms)} }
\startdata
38.3 && I && $64.49 \pm 5.41$ && $-11.24 \pm 0.02$ && $0.182 \pm 0.020$  \\  
38.5 && I && $67.34 \pm 3.89$ && $-11.25 \pm 0.02$ && $0.181 \pm 0.019$  \\ \hline 
38.3 && I-s && $16.62 \pm 1.37$ && $-10.99 \pm 0.04$ && $0.327 \pm 0.205$  \\ 
38.5 && I-s && $18.98 \pm 3.63$ && $-10.97 \pm 0.09$ && $0.356 \pm 0.219$  \\ \hline
38.3 && IIa && $100.51 \pm 8.54$\phantom{1} && $-10.59 \pm 0.02$ && $0.223 \pm 0.021$  \\ 
38.5 && IIa && $96.81 \pm 5.14$ && $-10.60 \pm 0.02$ && $0.228 \pm 0.020$  \\ \hline
38.3 && IIb && $78.33 \pm 5.98$ && $-10.75 \pm 0.02$ && $0.205 \pm 0.020$  \\
38.5 && IIb && $73.21 \pm 3.55$ && $-10.76 \pm 0.02$ && $0.202 \pm 0.019$  \\ \hline
38.3 && IIc && $35.67 \pm 5.22$ && $-10.84 \pm 0.04$ && $0.387 \pm 0.057$  \\
38.5 && IIc && $37.18 \pm 2.86$ && $-10.83 \pm 0.02$ && $0.403 \pm 0.037$  \\ \hline
\enddata
\tablenotetext{\rm \scriptsize a}{~The velocity linewidth was measured at FWHM.}
\end{deluxetable*}

Three Gaussian components were fitted to spectral feature~II at 38.3~GHz; we have designated them as IIa, IIb, and IIc. The intensity, center velocity, and FWHM linewidth of these three components are listed in Table~\ref{tline} and shown in Figure~\ref{fig:IF2wG}. The peak intensities of components IIa and IIb, 100.51 and 78.33~\jybeam\ respectively, are larger than that of component~I listed in Table~\ref{tline}, and their center velocities are $-10.59$ and $-10.75$~\kms. Like component~I, both component~IIa and IIb are narrow, with FWHM linewidths of 0.22~\kms\ and 0.21~\kms\ respectively. Meanwhile, component~IIc has lower intensity (35.67~\jybeam), broader FWHM linewidth (0.39~\kms), and a center velocity of $-10.84$~\kms.
The resultant Gaussian profile obtained by summing component~I and I-s, and separately summing IIa, IIb, and IIc is shown in Figure~\ref{fig:IF2}, together with the residuals from the fit. Again, the situation was similar at 38.5~GHz, where components IIa and IIb had higher intensity than component~I (see Table~\ref{tline}). Within the errors, the center velocities and linewidths of IIa and IIb at 38.5~GHz were the same as those at 38.3~GHz. Component~IIc also had lower intensity and broader linewidth than IIa and IIb at 38.5~GHz. In the approach that was followed, as described above, the two velocity features in both the 38.3 and 38.5\,GHz maser transitions were treated separately and independently when fitting Gaussian components to their profiles. This was necessitated by the simultaneous optimization of the Gaussian fits to the Stokes $I$ spectra and their scaled derivatives to the Stokes $V$ spectra. This process also took into account the minimization of the residuals while using as few Gaussian components as possible.

\begin{figure}[t!]
\epsscale{1.2}
\plotone{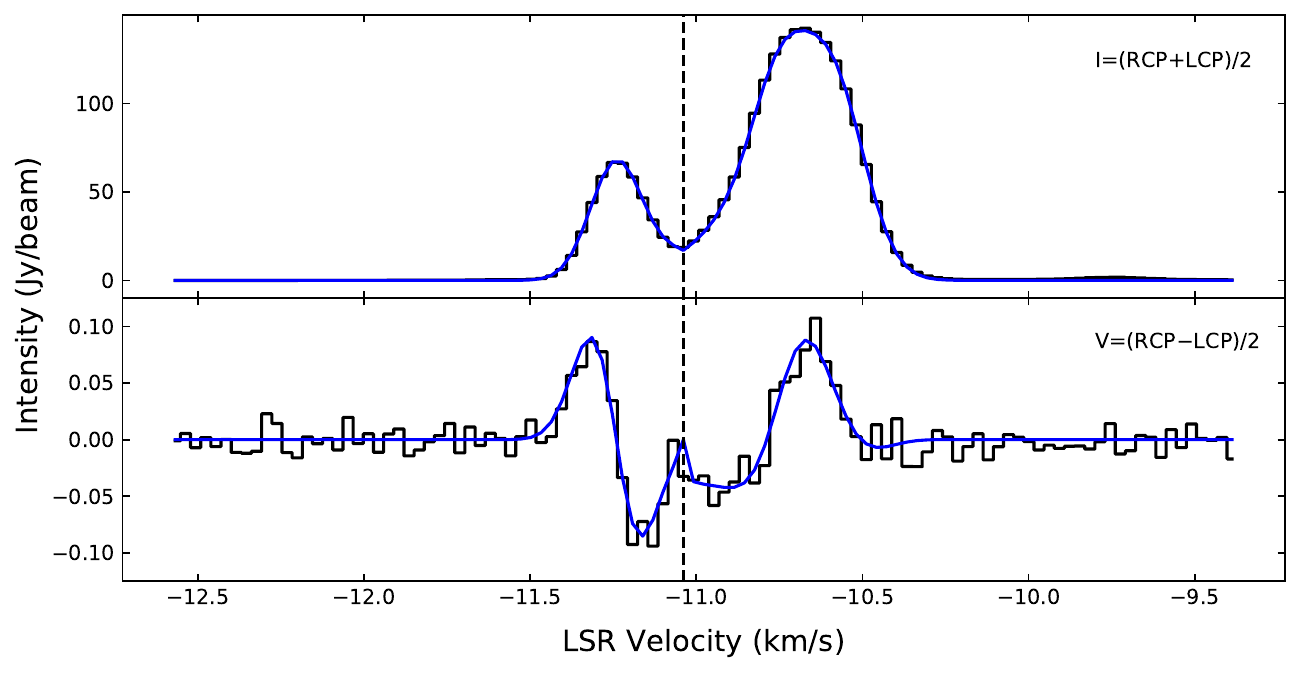}
\caption{Stokes $I$ (upper panel, black histogram-like line) and Stokes $V$ (lower panel, black histogram-like line) profiles toward NGC~6334\,F at 38.3~GHz. The blue curve in the upper panel is the same as that shown in, and described in the caption to, Figure~\ref{fig:IF2}. The blue curve superposed on the Stokes~$V$ profile in the lower panel is the sum of the solid red and dashed blue curves shown in the lower panel of Figure~\ref{fig:IF2zBcompL} to the left of the dashed vertical line, and the sum of the magenta, solid blue, and green curves in the lower panel of Figure~\ref{fig:IF2zBcompL} to the right of the dashed vertical line.  \label{fig:IF2zB} }
\end{figure}

\begin{figure}[t!]
\epsscale{1.2}
\plotone{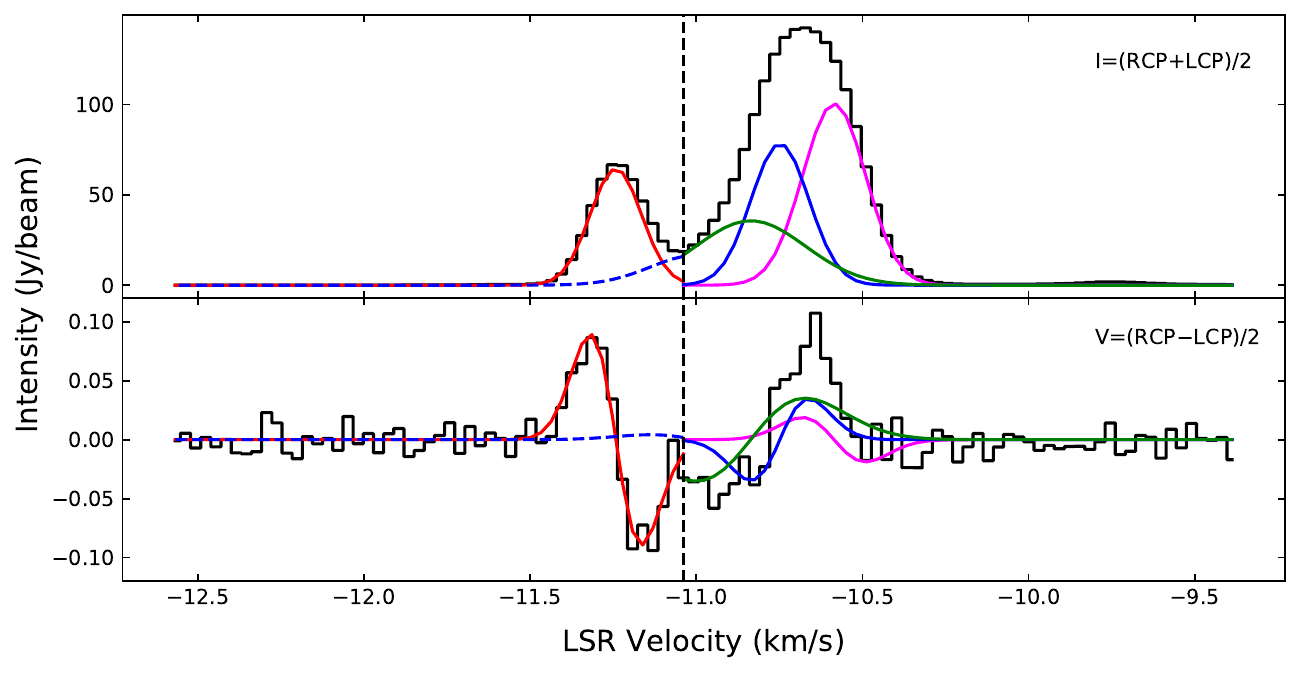}
\caption{Stokes $I$ (upper panel, black histogram-like line) and Stokes $V$ (lower panel, black histogram-like line) profiles toward NGC~6334\,F at 38.3~GHz, as in Figure~\ref{fig:IF2zB}. The solid red and dashed blue curves in the upper panel are Gaussian components I and I-s, respectively. The magenta, solid blue, and green curves are components IIa, IIb, and IIc, respectively. The fit parameters for all five components are listed in Table~\ref{tline}. The curves superposed on Stokes~$V$ in the lower panel are the derivatives of the corresponding colored curves in the upper panel, scaled by the fitted value of $z\Blos$ for each curve listed in Table~\ref{tzB}. \label{fig:IF2zBcompL} }
\end{figure}

\begin{figure}[t!]
\epsscale{1.2}
\plotone{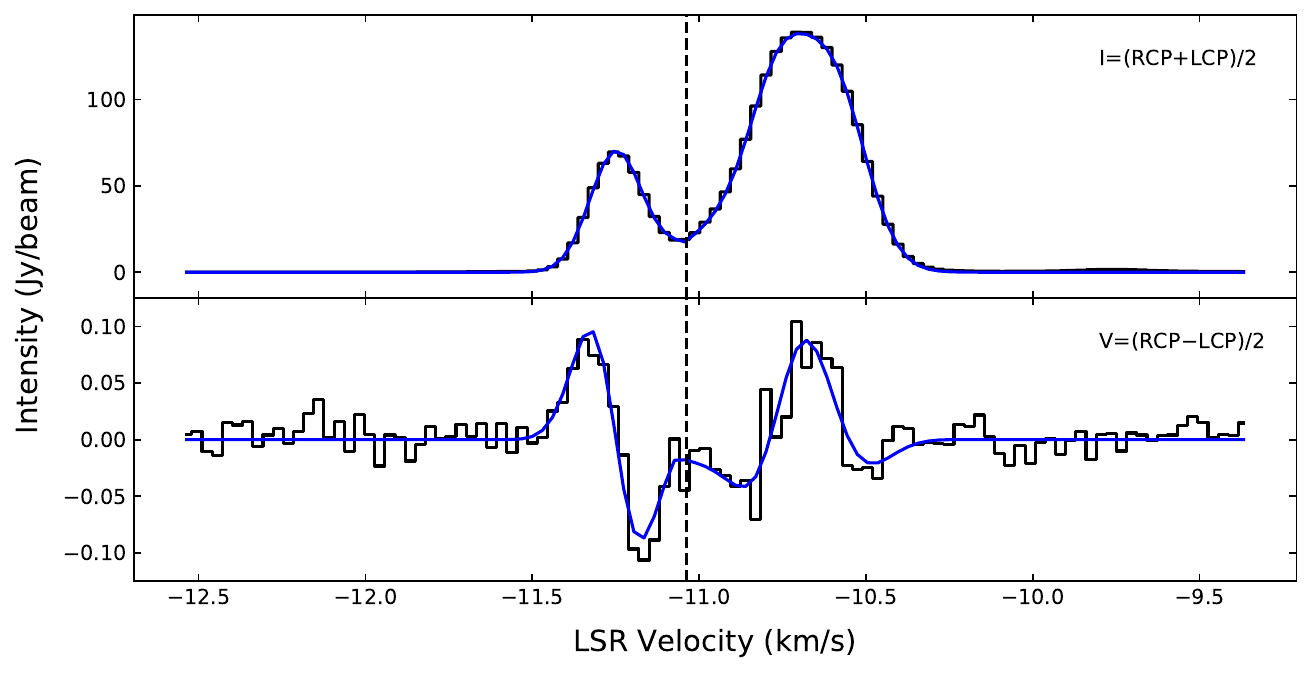}
\caption{As in Figure~\ref{fig:IF2zB}, but at 38.5~GHz. \label{fig:IF3zB} }
\end{figure}

\begin{figure}[t!]
\epsscale{1.2}
\plotone{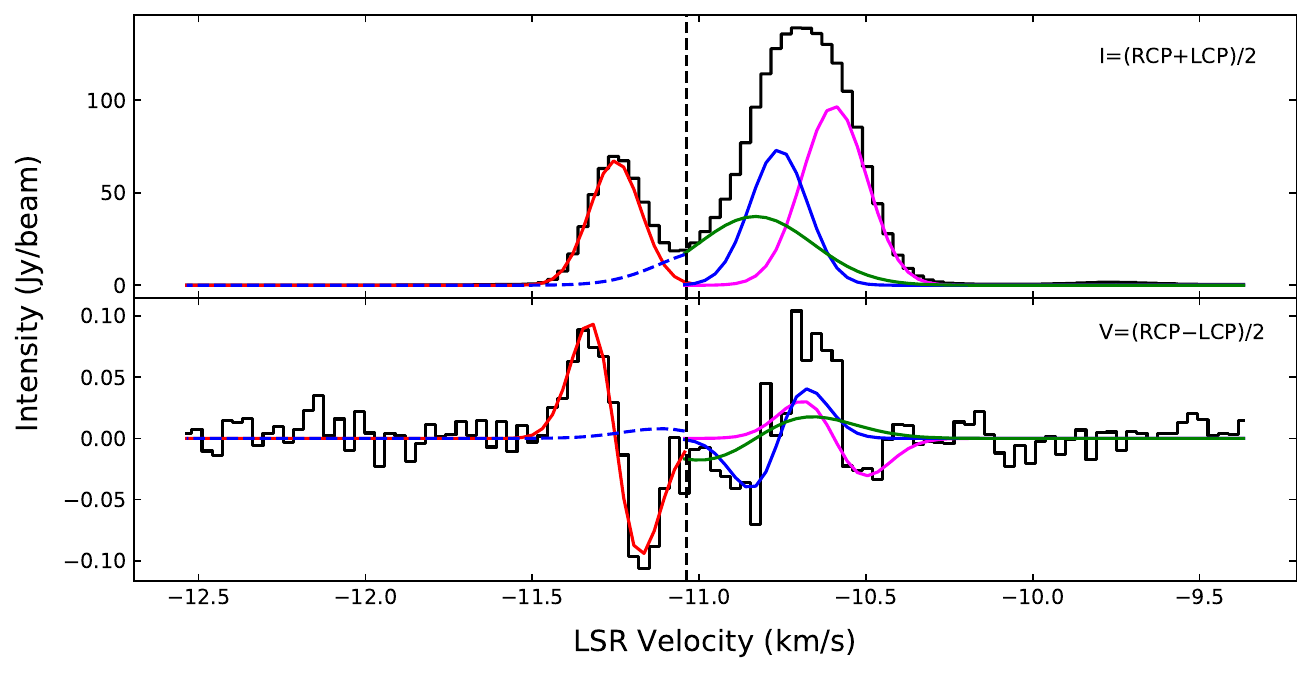}
\caption{As in Figure~\ref{fig:IF2zBcompL}, but at 38.5~GHz. \label{fig:IF3zBcompL} }
\end{figure}

The observed spectral profiles shown in Figure~\ref{fig:IF2} and \ref{fig:IF2wG} are the Stokes~$I$ profiles at 38.3~GHz. The Stokes $I$ and $V$ profiles toward NGC~6334\,F at 38.3~GHz are shown together in Figure~\ref{fig:IF2zB}. Following our usual procedure (see, e.g., \citealt{momjian+2017}), we fit the Stokes $V$ profile to the derivative of the Stokes $I$ profile and a scaled replica of the $I$ profile using the equation (\citealt{troland+1982}; \citealt{sault+1990}):
\begin{equation} 
V = aI + \frac{b}{2}\, \frac{dI}{d\nu}
\label{e.1} 
\end{equation}
The fit parameter $a$ is included to account for small calibration errors in RCP versus LCP and we obtained $a \lesssim 10^{-3}$. The fit parameter $b = z\Blos$, where $z$ is the Zeeman splitting factor and \Blos\ is the line-of-sight magnetic field strength. We fitted independently for spectral features I and II shown in Figure~\ref{fig:IF2}; the dashed vertical line in Figure~\ref{fig:IF2zB} separates the channels that we included in the fits for each of these two features. The fits were carried out using the AIPS task ZEMAN (\citealt{greisen+2017}), which allows for multiple Gaussian components in Stokes~$I$ to be fitted simultaneously to Stokes~$V$ for different values of $b$ as was done, for example, for the three components~IIa-IIc. Figure~\ref{fig:IF2zBcompL} shows the individual Gaussian components in the upper panel, and the derivatives of each component scaled by the fitted value of $z\Blos$ in the lower panel. Unlike in Figure~\ref{fig:IF2wG}, components~I and I-s are shown only to the left of the dashed vertical line, and components IIa-IIc are shown to the right of the dashed vertical line; that is, confined to the velocity space in which they were summed to obtain the blue profiles superposed on the black histogram-like profiles in the upper and lower panels of Figure~\ref{fig:IF2zB}. Figure~\ref{fig:IF3zB} is the equivalent of Figure~\ref{fig:IF2zB} and Figure~\ref{fig:IF3zBcompL} is the equivalent of Figure~\ref{fig:IF2zBcompL}, but at 38.5~GHz.

The results of the fitting of Stokes~$V$ using equation~(\ref{e.1}) are given in Table~\ref{tzB}. Although \citet{lankhaar+2018} published values of the Zeeman splitting factor for several \meth\ maser lines, values of $z$ for 38.3 and 38.5~GHz lines are not available, so we will leave our results in terms of $z\Blos$ in Hz. We obtained three significant detections and two upper limits. Generally, we consider a detection significant if the signal-to-noise ratio in $z\Blos$ is 3-$\sigma$ or higher. Here, we impose a stronger criterion; we will consider the detection significant only if the signal-to-noise ratio in $z\Blos$ is 3-$\sigma$ or higher at both 38.3~GHz and 38.5~GHz. Component~I listed in Table~\ref{tline} shows a significant detection of $-45.10 \pm 3.40$~Hz at 38.3~GHz, and $-46.08 \pm 3.63$~Hz at 38.5~GHz. By convention, a negative value for \Blos\ implies that the line-of-sight magnetic field is pointing toward the observer. Component~IIa and IIb also show significant detections, and \Blos\ traced by component~IIa points toward the observer, whereas \Blos\ traced by component~IIb points away from the observer. Component~IIa shows a significant detection of $-7.48 \pm 1.81$~Hz at 38.3~GHz, and $-12.88 \pm 2.55$~Hz at 38.5~GHz. Finally, component~IIb shows a significant detection of $16.26 \pm 2.59$~Hz at 38.3~GHz, and $20.07 \pm 3.96$~Hz at 38.5~GHz. Although a value of $z$ is not available for these two transitions, values published by \citet{lankhaar+2018} for many prominent \meth\ transitions are near 1~Hz mG$^{-1}$, and \Blos\ values of 8-46~mG appear reasonable in such regions.

\begin{deluxetable}{cccccccccccl}
\tablenum{3}
\tablewidth{0pt}
\tablecaption{Zeeman Effect Measurements
	\protect\label{tzB}}
\tablehead{
	\colhead{Frequency} && \colhead{} && \colhead{z\Blos} \\
	\colhead{(GHz)} && \colhead{Component} && \colhead{(Hz)}  }
\startdata
\multicolumn{5}{c}{\textbf{Significant Detections}} \\ \hline
38.3 && I && $-45.10 \pm 3.40$  \\ 
38.5 && I && $-46.08 \pm 3.63$  \\ \hline
38.3 && IIa && \hspace{0.08in}$-7.48 \pm 1.81$  \\ 
38.5 && IIa && $-12.88 \pm 2.55$  \\ \hline
38.3 && IIb && $\phantom{-}16.26 \pm 2.59$  \\ 
38.5 && IIb && $\phantom{-}20.07 \pm 3.96$  \\ \hline
\multicolumn{5}{c}{\textbf{Upper Limits}} \\ \hline
38.3 && I-s && $-15 \pm 49$ \\ 
38.5 && I-s && $-27 \pm 41$ \\ \hline
38.3 && IIc && $\phantom{-}68 \pm 10$ \\ 
38.5 && IIc && $\phantom{-}34 \pm 14$ \\ \hline
\enddata
\end{deluxetable}

\vspace{0.27in}

\section{Discussion} \label{sec:disc}

We observed the 38.3 and 38.5~GHz Class~II \meth\ maser transitions toward NGC~6334\,F with the VLA for the Zeeman effect. At both 38.3 and 38.5~GHz, the observed spectral profiles contain two prominent velocity features. We fitted one with a narrow Gaussian component and a shallow broad component, which we have labeled as components I and I-s respectively (Table~\ref{tline} and Figure~\ref{fig:IF2wG}). The other was fitted with two narrow Gaussian components labeled IIa and IIb, and a lower intensity broad component, which we have labeled as IIc. Our observed maser profiles are consistent with the Australia Telescope Compact Array (ATCA) observations of \citet{ellingsen+2018}. However, it is worth noting that their velocity resolution was 0.3~\kms, a factor of $\sim$10 coarser compared to ours. They observed our two prominent velocity features (which we have labeled as I and II in Figure~\ref{fig:IF2}) to be blended in velocity space, but with a significant asymmetry on the left side of the profile (near $-11$~\kms), consistent with the presence of our spectral feature~I. Of interest is that the single dish observations of \citet{haschick+1989} with 0.065~\kms\ velocity resolution show feature~II but only weak emission at the velocity corresponding to feature~I. This could be due to variability, but there is no way to directly compare the two observations taken with such different instruments so many years apart. The base of the spectral profile observed by \citet{haschick+1989} at both 38.3 and 38.5~GHz indicates the presence of a broad emission component, consistent with the shallower and broader components I-s and IIc that we have fitted (Figure~\ref{fig:IF2wG}). In both the single dish observations of \citet{haschick+1989} and the ATCA observations of \citet{ellingsen+2018}, the 38.5~GHz spectral profile has higher intensity than the 38.3~GHz profile, consistent with our VLA observations (Table~\ref{tline}).

We have obtained significant detections of the Zeeman effect in components I, IIa, and IIb (Table~\ref{tzB}), with $z\Blos$ in the range 8-46~Hz. Obtaining the line-of-sight magnetic field values from these is possible only if we know the value of the Zeeman splitting factor $z$ for the 38.3 and 38.5~GHz transitions. \citet{lankhaar+2018} published values of $z$ for a wide range of \meth\ maser transitions, but not for the 38~GHz transitions. If, however, the values of $z$ for these transitions are near 1~Hz~mG$^{-1}$, as it is for many prominent transitions of \meth, then we would get \Blos\ values of the same order as $z\Blos$ in these regions. Values in the range 8-46~mG appear reasonable for high mass star forming regions. For example, from the statistical analysis of a flux-limited sample of 6.7~GHz Class II \meth\ masers, \citet{surcis+2022} found line-of-sight magnetic fields in the range 9-40~mG. It is worth noting that the \meth\ maser transitions result from a series of hyperfine transitions, each of which has a different $z$. Therefore, the magnetic field values would be different depending on which hyperfine transition, or combination of hyperfine transitions, is responsible for the maser transition. Also, it is worth noting that the reversal in sign of \Blos\ from component IIa to IIb is usually interpreted by astronomers as a field reversal in the regions traced by these masers. However, \citet{lankhaar+2018} have proposed a different scenario in which the change in sign is caused by the population inversion of two different hyperfine transitions. Values in the range 8-46~mG are also in general agreement with field strengths of 4.4-6.4~mG measured in 1.6 and 6.0~GHz OH maser lines in NGC~6334\,F by \citet{chanapote+2019}.

%

Following standard procedure in reporting the Zeeman effect, we discuss now whether the detected signal could be caused by instrumental effects, or by processes other than the Zeeman effect. It is well known that a velocity gradient across an extended source could mimic a Stokes~$V$ signal caused by the Zeeman effect. However, masers are compact sources confined to a narrow velocity range. Therefore, it is unlikely that the observed Stokes~$V$ signal reported in this paper is due to such a velocity gradient. Moreover, the close agreement of the detected $z\Blos$ value at 38.3 and 38.5~GHz, i.e., two transitions observed in independent spectral windows, gives us confidence that this is not a spurious detection. Processes other than the Zeeman effect that could contribute to structure in the Stokes~$V$ profile include, for masers with strong linear polarization, changes in the orientation of the magnetic field along the line of sight that could cause rotation of the linear polarization vectors to produce circular polarization (\citealt{wiebe+1998}). In order to optimize our observations for detection of the Zeeman effect, we did not propose for full polarization observations. We note, however, that for a wide array of Class~II \meth\ maser transitions, \citet{breen+2019} found linearly polarized emission in the range 1.5-7.5\%. If the 38~GHz Class~II \meth\ maser transitions reported in this paper have similar levels of linear polarization, then it is unlikely that the observed Stokes~$V$ could be caused by the rotation of linear polarization vectors due to changes in the magnetic field orientation along the line of sight. Another possible non-Zeeman origin  comes from maser radiation scattering off foreground molecules that can enhance antisymmetric spectral profiles in Stokes~$V$ (\citealt{houde+2014}).
In such cases, if the Stokes~$V$ were considered to be caused solely by the Zeeman effect, then the resulting magnetic field strengths would be too large, as demonstrated for SiO masers by \citet{houde+2014}. Unless $z$ for 38~GHz transitions is determined to be extremely low so that the values of $z\Blos$ we have detected translate to very high magnetic fields, it appears unlikely that this effect is of consequence for our observations. Finally, if the maser stimulated emission rate $R$ is larger than the frequency shift due to the Zeeman effect $g\Omega$, which is the product of the Land\'e g-value for the upper state of the transition and the gyrofrequency, $\Omega$, of the electron, a rotation of the axis of symmetry for the molecular quantum states could also cause circular polarization (\citealt{vlemmings2011}). In our observations toward NGC~6334\,F, we have $g\Omega \approx 10$~s$^{-1}$.
The stimulated emission rate $R$, taken from \citet{vlemmings2011}, is 
\begin{equation} 
	R \simeq \frac{AkT_b\Delta\Omega}{4\pi h\nu} \label{e.2} 
\end{equation}
where $k$ and $h$ are the Boltzmann and Planck constants, respectively, $\nu$ is the frequency of the maser transition, $A$ is the Einstein coefficient, equal to $4.726 \times 10^{-8}$~s$^{-1}$ for the $6_2 \rightarrow 5_3$~A$^{-}$ transition at 38.3 GHz, and $4.829 \times 10^{-8}$~s$^{-1}$ for the $6_2 \rightarrow 5_3$~A$^{+}$ transition at 38.5 GHz, $T_b$ is the maser brightness temperature, and $\Delta \Omega/4\pi \approx 10^{-3}$ is a conservative estimate for the maser beaming angle (\citealt{nesterenok+2016}). The lower limit on $T_b$ from our observations is $10^6$~K. Using these values, equation~(\ref{e.2}) gives $R \approx 10^{-5}$~s$^{-1}$, which in turn implies that $R \ll g\Omega$.
Therefore it is unlikely that a rotation of the axis of symmetry is causing the splitting that results in the observed Stokes~$V$ profile. Moreover, such an effect would cause an intensity-dependent polarization, but component IIa with higher intensity than IIb has a lower $z\Blos$ than that detected in IIb.


Class~II \meth\ masers are exclusively found in high mass star forming regions and are located close to the protostar, unlike Class~I \meth\ masers which are located in outflows. However, there is as yet no consensus on the structures that they trace in these protostellar environments. One scenario puts them in the accretion disk itself, whereas an alternative places them in the intermediate region between the outflow and the disk (\citealt{sanna+2010}; \citealt{goddi+2011}; \citealt{sugiyama+2014}; \citealt{bart+2020}). We can use our calculated value for \Blos\ to compare the magnetic energy density with other relevant quantities in NGC~6334\,F. The magnetic energy density is given by $B^2/8\pi$, where $B^2 = 3\Blos^2$, as determined by \citet{crutcher+1999}  on statistical grounds; we note that this is strictly only valid for an ensemble of measurements. 
For \Blos=8~mG reported in Section~\ref{sec:res}, which is the lowest magnetic field value we measure, the magnetic energy density is equal to $7.6 \times 10^{-6}$~erg~cm$^{-3}$. The kinetic energy density is a relevant quantity with which to compare this value; it is given by $(3/2)\, m n \sigma^2$, where $m = 2.8~m_p$, with $m_p$ being the proton mass, the numerical factor of 2.8 also accounting for 10\% helium. The particle density is $n=10^6$~cm$^{-3}$, from modeling of several Class~II \meth\ maser lines in NGC~6334\,F, including the 38~GHz transitions (\citealt{cragg+2001}). The velocity dispersion is given by $\sigma = \Delta v/(8 \ln 2)^{1/2}$. Rather than using the narrower $\Delta v$ from maser observations, which may not be indicative of the thermal motions in the clump of gas where the maser transition arises, we use the broader value of 3~\kms, from the velocity gradient in HC$_3$N and NH$_3$ emission lines toward NGC~6334\,F (\citealt{jackson+1988}; \citealt{bachiller+1990}). This gives a kinetic energy density equal to $1.1 \times 10^{-7}$ ergs~cm$^{-3}$. This means that the magnetic energy density in the region traced by the masers in NGC~6334\,F is at least comparable to, if not larger than, the kinetic energy density. If the 38~GHz \meth\ masers occur in a rotating accretion disk, another relevant quantity of interest is the rotational energy density, given by $(1/2)\, I \omega^2/V$, where $\omega$ is the angular velocity and $V$ is the volume of the disk. The moment of inertia $I$ is of the order $MR^2$; $M$ is the mass and $R$ is the radius of the disk. Using a rotational velocity of $v=4$~\kms\ for methanol maser disks from \citet{norris+1998}, where $v = R\omega$, the expression for the rotational energy density then becomes $(1/2)\, (M/V)\, v^2$, where $M/V = 2.8\, m_p\, n$, with $m_p$ being the proton mass, and $n$ the particle density with a value of $10^6$~cm$^{-3}$, as discussed above. The derived rotational energy density is then about $3.7 \times 10^{-7}$ ergs~cm$^{-3}$. Once again, the magnetic energy density is at least comparable to, if not greater than, the rotational energy density. Therefore, if the 38~GHz Class~II \meth\ masers are located in the accretion disk around the protostar, then the magnetic field likely plays a significant role in shaping the dynamics in that accretion disk.

\section{Conclusion} \label{sec:conc}

We observed the 38.3 and 38.5~GHz Class~II \meth\ maser transitions toward the high mass star forming region NGC~6334\,F for the Zeeman effect. Both transitions have similar spectral profiles, each with two prominent spectral features. We fitted one of these two features with a single narrow Gaussian and a shallow broad component, with FWHM linewidths of 0.18~\kms\ and 0.33~\kms\ respectively at 38.3~GHz. These components were labeled I and I-s respectively. We fitted the other prominent spectral feature with two narrow Gaussian components labeled IIa and IIb, and a broad, lower intensity component IIc; at 38.5~GHz, component~IIc with a FWHM linewidth of 0.403~\kms\ is almost twice as wide as component IIb, which has a FWHM linewidth of 0.202~\kms. The center velocities of these components are quite close, ranging from $-10.59$~\kms\ to $-11.25$~\kms.


We have obtained significant detections of the Zeeman effect in components I, IIa, and IIb. Values of $z\Blos$ in these masers range from 8 to 46~Hz, and there is a change in sign from component IIa to IIb. The Zeeman splitting factor $z$ for these 38~GHz transitions is not known, but if it is of the order $\sim$1~Hz mG$^{-1}$ as it is for a range of \meth\ maser transitions, then the magnetic fields in the regions traced by these masers would be in the range 8-46~mG. There is a reversal in the sign of $z\Blos$ from component IIa to IIb, usually interpreted as a reversal in \Blos\ from one component to another; alternatively, it may be a result of different hyperfine transitions being responsible for the maser transitions. Magnetic fields in the range 8-46~mG agree well with fields detected in the better-known 6.7~GHz \meth\ transition. Class~II \meth\ masers are known to form close to the protostar in accretion disks or in the interface region between the accretion disk and outflow. Using the lowest magnetic field value of \Blos = 8~mG from Table~\ref{tzB}, we find that the magnetic energy density is at least comparable to, if not greater than, the kinetic energy density and the rotational energy density in the disk. Such fields may exert significant influence on the dynamics of the accretion disk and likely play an important role in the star formation process.



\acknowledgments 
We thank an anonymous referee for insightful comments that helped to improve the paper. APS acknowledges a Summer Research Grant from the University Research Council (URC) at DePaul University.


\vspace{5mm}
\facilities{VLA}
\software{AIPS \citep{greisen+1990}}

\clearpage

\end{document}